\documentclass[twocolumn,showpacs,nofootinbib,preprintnumbers,amsmath,amssymb]{revtex4}
\usepackage{graphicx}
\usepackage{dcolumn}
\usepackage{bm}
\newcommand{\be}{\begin{equation}}
\newcommand{\ee}{\end{equation}}
\newcommand{\bef}{\begin{figure}}
\newcommand{\eef}{\end{figure}}
\newcommand{\bea}{\begin{eqnarray}}
\newcommand{\eea}{\end{eqnarray}}

\newcommand{\by}{{\bf y}}
\begin{document}

\title{Gravitational force in an infinite one-dimensional
Poisson distribution}

\author{A. Gabrielli$^{1,2}$ and M. Joyce$^{3,4}$} 
\affiliation{$^1$SMC, CNR-INFM, Physics Department, University 
``Sapienza'' of Rome, Piazzale Aldo Moro 2, 00185-Rome, Italy}
\affiliation{$^2$Istituto dei Sistemi Complessi - CNR, Via dei Taurini 19, 
00185-Rome, Italy}
\affiliation{$^3$Laboratoire de Physique Nucl\'eaire et Hautes \'Energies,\\
Universit\'e Pierre et Marie Curie - Paris 6,
CNRS IN2P3 UMR 7585, 4 Place Jussieu, 75752 Paris Cedex 05, France}
\affiliation{$^4$Laboratoire de Physique Th\'eorique de la Mati\`ere Condens\'ee,\\
Universit\'e Pierre et Marie Curie - Paris 6,
CNRS UMR 7600, 4 Place Jussieu, 75752 Paris Cedex 05, France}

\hyphenation{Lan-ge-vin}
\begin{abstract}
We consider the statistical properties of the gravitational field $F$
in an infinite one-dimensional homogeneous Poisson distribution of particles,
using an exponential cut-off of the pair interaction to control and 
study the divergences which arise. 
Deriving an exact
analytic expression for the probability density function (PDF) $P(F)$,
we show that it is badly defined in the limit in which 
the well known Holtzmark distribution is obtained in   
the analogous three-dimensional case. A well defined $P(F)$ may, however, be obtained 
in the infinite range limit by an appropriate renormalization of the 
coupling strength, giving a Gaussian form. Calculating the spatial correlation
properties we show that this latter procedure has a trivial physical
meaning. Finally we calculate the PDF and correlation properties of
{\it differences of forces} (at separate spatial points), which are
well defined without any renormalization. We explain that the convergence 
of these quantities is in fact sufficient to allow a physically meaningful
infinite system limit to be defined for the clustering
dynamics from Poissonian initial conditions.
\end{abstract}

\pacs{02.50.-r,05.40.-a,61.43.-j}
\maketitle


\section{Introduction}

In attempting to understand better the complex dynamics of classical
non-relativistic self-gravitating systems relevant in astrophysics
and cosmology, it is of interest to study simplified toy models
in one dimension. 
For systems of a finite number of particles, the evident
such model consists of particles on a line interacting 
by forces independent of  their separation (or, equivalently,
infinite parallel sheets embedded in three dimensions). This so-called
``sheet model''  has been quite extensively studied in the
literature (see, e.g., \cite{sheet1, sheet2, sheet3, severne+luwel,          
rouet+feix, tsuchiya+gouda, miller_1dreview} and
references therein) for its interest both as a 
toy model for gravity in three dimensions and, more generally, as a 
toy model for systems with long-range interactions.
In the study of the formation of structure in the universe
in cosmology, the problem may be well approximated, over a 
large range of length and time scales, by a variant of the 
simple three-dimensional (3D) Newtonian problem: particles 
belonging to an {\em infinite} distribution of particles 
evolve under equations of motion which are formally
identical to those for a finite system, up to simple 
modifications which take into account the expansion 
of the universe (see e.g. \cite{bagla_review, SL_2007}).
A few groups of
authors \cite{rouet_etal, yano+gouda, tatekawa+maeda, 
aurell_etal, miller+rouet_2002, aurell+fanelli_2002b, 
miller+rouet_2006, valageasOSC_1, 
valageasOSC_2, miller_etal_2007} have 
proposed different variants  of the simple
one-dimensional (1D) sheet model to mimic these equations. We have underlined
in a recent paper \cite{sl1d} that a fundamental 
question about any such model, just as in three dimensions, 
is whether the gravitational force term --- which
is simply the infinite sum representing the force 
exerted by all other particles on the given 
particle --- is well defined in the class of infinite 
distributions one wishes to study (which will represent 
the initial conditions for structure formation in 
cosmology). In \cite{sl1d} we have rigorously shown 
that, with an appropriate prescription, the infinite sum 
in one dimension results in a finite and simple
expression for the gravitational force acting on particles, 
in a specific class of infinite point distributions: 
infinite lattices subjected to a class of stochastic 
perturbations. In this paper we consider this issue of
the definedness of the 1D gravitational force in an 
infinite and homogeneous Poisson particle distribution, and 
then the related 
(but not, as we will discuss, identical) 
question of 
whether an infinite system limit for the dynamics of 
gravitational clustering can be usefully defined in
this case.

The approach we use to the question of the definedness of the
force follows the seminal work of Chandrasekhar dating back 
to 1943 \cite{chandra}: the infinite
point distribution is described by a stochastic point process in
infinite space, and one considers the statistical properties
of the gravitational field at an arbitrary point which is
itself then a stochastic quantity.  This approach has been 
adopted and generalized in various cases by ourselves and 
our collaborators as well as other authors, in the treatment
of both gravitational and other forces \cite{force1,force2,force3,
force4,libro,delpopolo,dipole-poisson}. In his original
work Chandrasekhar derived the result for the PDF of the total 
gravitational field generated by a 3D infinite homogeneous
Poisson distribution,  which is given by the 
so-called {\em Holtzmark distribution}. It 
is the generalization of this case to one dimension which is the
subject of the present paper. We note that this question
has been treated also in a recent paper \cite{chavanis}. The 
results we present here extend
considerably, and allow us to explain the physical
meaning of, those presented (for one dimension) in this latter
paper. 

We recall at the outset that to define the gravitational force in any
dimension in {\it any} spatially homogeneous infinite mass
distribution with non-zero mean density it is necessary (but not
necessarily sufficient) to give a prescription for its calculation. 
This is the case because the gravitational pair force in $d$ dimensions,
which decays as $r^{1-d}$ with the separation $r$, gives  
a force on any particle due to the non-zero mean density 
which is badly
defined. In three dimensions the appropriate prescription (for the cosmological
problem) consists in the so-called ``Jeans' swindle'' (see,
e.g.,\cite{jeans, binney}): the contribution of the mean density is removed
and only density fluctuations are taken as the source for the
gravitational field. As discussed in \cite{kiessling} this ``swindle''
is a (well-defined) mathematical regularization of the gravitational
force acting on particles in the infinite system limit (at constant
non-zero mass density), which can be most simply re-stated as the
prescription that the force on each particle be calculated by summing
symmetrically about it (e.g. in spheres centred on the point).  In his
calculation of the PDF in three dimensions Chandrasekhar adopted, albeit without
explicit discussion, the prescription of summation in such spheres, and
other calculations in three dimensions (for other distributions) have done the
same. In \cite{force3} one of has shown that the adaption of this
prescription to one dimension, i.e.  summation in a symmetric interval about
each point, gives a well defined force PDF, in a class of infinite
perturbed lattices, only for pair forces which decrease with
separation, but not for the (separation independent) 1D
gravitational force. In \cite{sl1d} we have shown, however, that by
adapting a smooth version of this prescription --- calculating the
total gravitational field as the limit of the sum for an exponentially
screened gravitational pair interaction when the screening vanishes
continuously--- we obtain a well defined PDF also for the 
gravitational interaction. Thus while in three dimensions the Jeans' swindle can be
  implemented in various different ways with the same finite results
  in a wide class of stochastic point-mass distributions, the use of a
  smooth version of it (rather than a sharp ``top-hat'' version) has
  to be preferred in one dimension. This is due to the fact that \cite{sl1d}
  even in a wide class of very uniform stochastic point process,
  stochastically perturbed lattices, a top-hat prescription leaves
  fluctuating and non-converging (i.e. undetermined) boundary
  contributions which are erased by a smooth
  regularization. It is this formulation of the Jeans' swindle which
we will employ here, although one can, as we will discuss, in fact
equally use the ``top-hat'' prescription and obtain equivalent results
in the specific case of a Poisson distribution.

The first result of this paper is an
exact expression for the PDF of the screened 1D
gravitational force in an infinite Poisson 
distribution, derived in Sec.~\ref{regu}. 
More specifically we give the exact
expression for the cumulants of the PDF. Using this
expression we show that the PDF is, as can be 
anticipated from a simple qualitative analysis 
we give in Sec.~\ref{def}, badly defined in the
limit that the inverse screening length $\mu$
goes to zero. This in three dimensions corresponds to the limit
in which instead the well-defined Holtzmark distribution 
has been obtained by Chandrasekhar. 

In Sec.~\ref{ren} we discuss {\it two} 
other ways in which the infinite range limit 
may be taken, both involving a renormalization
of the other system parameters (mean density and coupling strength),
and give the (different) asymptotic form of the PDF 
obtained in each case. 
The meaning of these 
renormalisation procedures is explained in
Sec.~\ref{corr} by means of an exact calculation,
again for the screened interaction, of the 
force-force correlation function. This shows
that the non-trivial renormalisation procedures
actually give a {\it spatially uniform} force (acceleration) field,
as in the corresponding limit only the (initially)
divergent contribution to the force from particles
infinitely far away survives. This means that 
the dynamics under the renormalised forces 
corresponds to a trivial (albeit stochastic) rigid translation 
of all particles, while the spatially varying 
component of the forces, which would lead
to non-trivial relative motions (i.e. in this
case, clustering) vanishes.

This discussion leads us naturally to focus on 
the fact that the spatially non-trivial part 
of the correlation function is in fact well
defined in the simple (unrenormalized) infinite
range limit. We show in Sec.~\ref{rel-field} 
that this is reflected more generally in the
fact that the {\it differences of forces} between 
points at some fixed distance is manifestly 
well defined and independent of the distribution
outside the interval they enclose. We derive
an exact expression for the PDF (specifically,
again, for its cumulants) of the difference in
the (unscreened) gravitational force between
two points in space, both without and with 
the Jeans' swindle (implemented either with
a top-hat or the screening prescription). 
As we are interested in the {\it clustering dynamics} 
manifested in the finite system, i.e. {\it relative motion of
particle initially contained in any finite 
region}, we discuss whether the definition
of the relative force PDF is sufficient
to make the infinite volume limit
meaningful. We argue that, if the Jeans' 
swindle is employed, the limit can indeed 
be defined.

\section{1D Gravitational Field in a Poisson distribution}
\label{def}

Let us consider a homogeneous  Poisson particle distribution 
(i.e. a random particle distribution) \cite{libro,poisson} 
on the interval $[-L,L']$ with average density $n_0$.
In other words it is characterized by a microscopic density
\be
n(x)=\sum_{i=1}^N \delta(x-x_i)
\label{n}
\ee
where $x_i$ is the position of the $i^{th}$ particle, and such that 
\bea
&&\left<n(x)\right>=\lim_{L,L'\to \infty} \frac{N}{L+L'}=n_0>0\nonumber\\
&&\left<n(x)n(x')\right>=n_0^2+n_0\delta(x-x')\,,
\label{n-cor}
\eea
where $\left<\cdot\right>$ means the usual ensemble average.
The second relation says that there is no correlation between the 
positions of different particles. More generally the joint probability 
density function (PDF) of the positions of all the particles is simply
\be
p_N(x_1,...,x_N)=\left(\frac{1}{L+L'}\right)^N
\label{p-x-n}
\ee 
The 1D version of the gravitational field generated at $y$ by a particle 
at $x$ may be written
\be 
f(x-y)=g\frac{x-y}{|x-y|}=g\cdot\mbox{sgn}(x-y)\,.
\label{f}
\ee   
The coupling coefficient $g$
gives the intensity of the interaction\footnote{It is equal to $2\pi G
  \Sigma$ when we derive the model from infinite parallel sheets in
three dimensions, where $G$ is Newton's constant and $\Sigma$ is the surface
  mass density of the sheets.}.
The total field at the point $y$ is therefore given by
\be
F(y)=\sum_{i=1}^N f(x_i-y)=\int_{-L}^{L'}dx\, n(x)f(x-y)\,.
\label{F}
\ee 
We want to study the statistical properties of this quantity in
the thermodynamic limit $N,L,L'\to\infty$ with the sole constraint
$N/(L+L')\to n_0$.  It is clear that in this limit the integral
Eq.~(\ref{F}) is, just as the analogous one in 3D gravity, 
ill defined  as $n(x)f(x-y)$ is not integrable at 
large $|x|$ in almost any realization of the Poisson particle 
system.  

It is simple to show by the following simple handwaving argument the
result we will find rigorously in the next section: while in 3D the
usual ``Jeans' swindle'' suffices to make the typical force in a
Poisson distribution well defined in the infinite system limit, this
is not the case in one dimension.  We recall that Jeans' swindle consists in
considering as the source of the field not the density field $n(x)$,
but the density fluctuations $\delta n(x)=[n(x)-n_0]$ from the
mean. For any central force, this is equivalent to considering the
whole $n(x)$ as the source, but summing symmetrically with respect to
the point $y$ where the field is evaluated, e.g. in spheres centered
on this point. In the introduction of Chandrasekhar's celebrated
derivation \cite{chandra} of the PDF of the total gravitational field
in a 3D homogeneous Poisson particle system, it is this latter
procedure which is implicitly followed. In one dimension this corresponds to
taking\footnote{This  prescription is adopted also in \cite{chavanis}.}
$L'=L+2y$ in Eq.~(\ref{F}). To see the difference between the two
cases (in one and three dimensions), it suffices to estimate, once the Jeans'
swindle is adopted, the order of contribution $\Delta F_{>R}$ to
Eq.~(\ref{F}) coming from sources at a distance greater than $R$ from
the point $\by$. In doing this we approximate the integral in
Eq.~(\ref{F}) with a sum over shells defined by the the sequence of
radii $R_n=2^nR$ with integer $n=0,1,2, ...$, i.e., radii equally
spaced on a logarithmic scale.  It is known that the typical density
fluctuation $\delta n_V$ in a given volume $V$ is, for a homogeneous
Poisson point process in any spatial dimension, of order $V^{-1/2}$. In
particular the volume of the $(n+1)^{th}$ shell is $V_n=A(2^{n}R)^d$
where $A$ is a geometrical prefactor depending on $d$ but not on
$n$. Therefore we can say that the typical density fluctuation in the
$(n+1)^{th}$ shell is $\delta n(R_n)\simeq
[A(2^{n}R)^d]^{-1/2}$. Given that the pair force
between two particles at distance $r$ is of order $r^{-d+1}$ in $d$
dimensions, it follows  that at sufficiently large $R$ we can
approximate Eq.~(\ref{F}) by
\be
\Delta F_{>R}\sim \sum_{n=0}^\infty A(2^{n}R)^d
\frac{[A(2^{n}R)^d]^{-1/2}}{(2^nR)^{d-1}}\sim C(d)R^{-d/2+1}\,,
\label{F-approx}
\ee 
with $C$ a positive constant depending on $d$. Let us start by analyzing 
Eq.~(\ref{F-approx}) for $d=3$.
We can simply verify that $C(d)$ is finite and therefore such a
contribution to the total force is finite for any $R$ and vanishes
for $R\to\infty$.
Thus, when one sums in spheres
about a given point, the typical force on a particle 
converges. This is the fundamental reason why, once the
Jeans' swindle in this form is adapted, the PDF of the
gravitational force is well defined (and given, as
derived in \cite{chandra}, by the so-called 
Holtzmark distribution). On the other hand, in $d=1$ one can see 
that $C(d)=+\infty$, i.e., 
at any finite $R$ the quantity $\Delta F_{>R}$ is divergent.
Moreover the $R$ dependence is pathologically increasing with $R$.
This means that, even if
one sums symmetrically, the contribution from {\em fluctuations}
around the mean density due to far away regions always give the 
dominant diverging contribution to Eq.~(\ref{F}).  
In the following we will demonstrate this result more formally, showing that
the PDF of the total field $F$ calculated in a symmetric window of size $L$
becomes broader and broader as $L$ increases, vanishing for any finite 
value of $F$ in the limit $L \to \infty$,  i.e., the force $F$ is an 
ill defined and completely undetermined stochastic quantity even in
this symmetric limit.

\section{PDF of the regularized force}
\label{regu}

In order to study in a controlled manner the statistical properties of
the 1D gravitational force which, as anticipated, may
be badly defined, we follow a procedure like that adopted often
in the context, notably, of quantum field theory: we introduce
a regularization and then study the behaviour of relevant 
physical quantities in the limit that this regularization 
is removed. The Jeans' swindle itself can, as discussed in 
\cite{kiessling}, be considered to be such a scheme: in the
usual ``top-hat'' implementation the regularization parameter
would then be the size of the symmetric interval (or radius of the
sphere in three dimensions) in which one sums. One can equally 
consider a smooth version in which the symmetric sum is
implemented by screening symmetrically the interaction.
As discussed in the introduction, it is the latter 
procedure we adopt here, as we have shown in 
\cite{sl1d} that this form is, in one dimension, actually 
preferable to the top-hat form because it gives, in a class
of more uniform distributions than those considered
here, a well defined (and physically meaningful) finite
force where the top-hat regularization does not.
We thus consider the pair interaction exerted by a particle at $x$ 
on another at $y$
\be
f_\mu(x-y)=g[\mbox{sgn}(x-y)]e^{-\mu|x-y|}\,,
\label{f-mu}
\ee 
introducing a cut-off length $\mu^{-1}$ characterising
an exponential screening of the ``bare'' gravitational 
interaction. We will take the limit $\mu\to 0^+$ at the end of 
our calculation of physically relevant quantities, and 
specifically, in the next section, to find the existence 
conditions for the PDF of the total force in this limit.

Given the distribution of $N$ particles in the interval $[-L,L']$
defined by Eqs.~(\ref{n}), the total field at $y$ is 
\be F_\mu(y)=\sum_{i=1}^N f_\mu(x_i-y)=
\int_{-L}^{L'}dx\,n(x)f_\mu(x-y)\,.
\label{F-mu}
\ee 
Without loss of generality let us now fix $y=0$.  Using
Eq.~(\ref{p-x-n}), the PDF of $F_{\mu}=F_{\mu}(0)$ given $N,L$ and
$L'$ can be written as\footnote{Note that, as in a Poisson system
  there is no correlation of the particle positions, the PDF of the
  total field at a spatial point is the same whether this point is
  assumed to be occupied or not. This is a feature specific to Poisson
  systems, which greatly simply the calculation of the force PDF
  compared to other cases (see, e.g., discussion in \cite{libro}).}
\be P(F_\mu;N)=\int..\int_{-L'}^L\prod_{i=1}^N \left[\frac{dx_i}{L+L'}
  \right]\delta\left[F_\mu-\sum_{i=1}^N f_\mu(x_i)\right]\,.
\label{P-F-N}
\ee
By using the identity
\be
\delta(z)=\int_{-\infty}^{+\infty}\frac{dq}{2\pi}\,e^{iqz}
\label{identity}
\ee
we can write
\bea
\label{P-F-N2}
&&P(F_\mu;N)=\int_{-\infty}^{+\infty}\frac{dq}{2\pi}\,e^{iqF_\mu}\times\\
&&\left(\int_{-L'}^L\frac{dx}{L+L'}
\,\exp\left\{-igq[\mbox{sgn}(x)]e^{-\mu|x|}\right\}\right)^N\nonumber\,.
\eea
In order to take the limit $(N,L,L')\to\infty$ with $N/(L+L')=n_0>0$,
we start by writing 
\bea
&&\int_{-L'}^L\frac{dx}{L+L'}\,\exp\left\{-igq[\mbox{sgn}(x)]
e^{-\mu|x|}\right\}
\nonumber\\ &&\equiv
1-\int_{-L'}^L\frac{dx}{L+L'}\left(1-\exp\left\{-igq[\mbox{sgn}(x)]
e^{-\mu|x|}\right\}\right)\,.\nonumber
\eea 
Since for $\mu>0$ we have
\[\left|\int_{-\infty}^{+\infty} dx\left(1-\exp\left\{-igq[\mbox{sgn}(x)]
e^{-\mu|x|}\right\}\right)\right|<+\infty\,,\]
the above limit of Eq.~(\ref{P-F-N2}) yields 
\[
P(F_\mu)=\int_{-\infty}^{+\infty}\frac{dq}{2\pi}\,
e^{iqF_\mu+{\cal G}(q;n_0,g,\mu)}\,.
\]
where we have called $P(F_\mu)$ the asymptotic shape of $P(F_\mu;N)$ for
$N,L,L'\to\infty$ with $N/(L+L')=n_0$ and 
\be 
{\cal G}(q;n_0,g,\mu)=-2n_0\int_0^\infty dx \left[1-\cos\left(qge^{-\mu
    x}\right)\right]
\label{P-F-N3}
\ee
is the {\em cumulant generating function} of the stochastic force $F_\mu$. 
The cumulant of $l^{th}$ order of $F_\mu$ is given by
\be
C_l(n_0,g,\mu)=i^l\left.\frac{d^l{\cal G}(q;n_0,g,\mu)}{dq^l}
\right|_{q=0}\;\;\mbox{for }l\ge 1 \,.
\label{cumu}
\ee
Note that 
$C_1=\left<F_\mu\right>$ and $C_2=\left<F_\mu^2\right>-
\left<F_\mu\right>^2$, and that for a Gaussian 
variable ${\cal G}$ is a quadratic function of $q$.
In our case we have that all the odd cumulants vanish 
by symmetry, while the
even ones are
\begin{equation}
\label{PDF-cumulants}
C_{2l}(n_0,g,\mu)=\frac{n_0g^{2l}}{\mu l}\,.
\end{equation}

\section{Renormalization schemes}
\label{ren}

Let us now analyze the behavior for $\mu\rightarrow 0$, i.e., when 
the screening length diverges. This is equivalent to applying the 
Jeans' swindle in the usual way. First of all we see that,
as anticipated in the previous section, all the non-zero
cumulants diverge ($\sim 1/\mu$) and thus the stochastic force 
$F$ becomes ill defined in the following sense: its PDF $P(F)$ at smaller and smaller $\mu$ 
becomes broader and broader vanishing point-wise at all $F$. 
This is analogous to 
the behavior of the PDF of the sum of $N$ identically distributed and independent random variables
with finite variance and zero mean: it becomes broader and broader 
as $N$ increases
and vanishes point-wise at all finite values for $N\to\infty$. 
We underline, that this limit $\mu\to 0$ is also analogous 
to the one behind Chandrasekhar's calculation \cite{chandra} in $d=3$
leading to the Holtzmark distribution; and likewise that
considered in obtaining a well defined PDF for the 1D
case for a class of perturbed lattices in \cite{sl1d}.

It is clear from Eq.~(\ref{PDF-cumulants}) that it is possible,
however, to converge to a well defined PDF in the limit 
$\mu \to 0$ if one appropriately renormalizes also the 
characteristic scales of the system. There are 
essentially two different possibilities:
\begin{itemize}
\item $n_0,\mu\to 0$ with $n_0/\mu=a>0$ and fixed $g>0$:
in this case all the cumulants, and therefore the PDF $P(F)$
are well defined with
\[C_{2l}=a\frac{g^{2l}}{l}>0\;\;\forall l\ge 1\,.\]
This limit is non Gaussian as even cumulants of order larger than two
are strictly positive. We call this case the {\em sparse limit} (SL)
as the particle density vanishes together with $\mu$.
\item $g,\mu\to 0$ with $g^2/\mu=b>0$ and fixed $n_0>0$:
in this case also $P(F)$ is well defined. However now $F$ becomes 
a Gaussian variable for which
\[C_2=b\frac{n_0}{2}\;\;\mbox{and }\,C_{2l}=0\;\mbox{for }l\ge 2\,.\]
In other words in this limit we have
\[P(F)=\frac{1}{\sqrt{\pi bn_0}}e^{-F^2/(bn_0)}\,.\]
This limit corresponds to the limit in which a
sort of {\em central limit theorem} applies to the force $F$. 
We will refer to it as 
the {\em weak interaction limit} (WIL). This is 
the limit which is considered in \cite{chavanis}.
\end{itemize}

Thus while the ``bare'' stochastic force $F$ as defined by
the simple sum (\ref{F}) is ill defined in the {\em infinite
volume} + {\em infinite force range} limit, the 
two ``renormalized'' forces $\phi=F\sqrt{\mu/n_0}$ 
and $\psi=F\sqrt{\mu}/g$ are well defined stochastic 
variables in the same limit.  However only the
latter is Gaussian.

The first limit is manifestly quite trivial: in the system
we consider there are just two characteristic length
scales $\lambda_1=1/n_0$ (typical interparticle distance) and
$\lambda_2=1/\mu$ (range of the interaction). Therefore a 
system in which these two lengths are substituted by 
two proportional ones
$\lambda_1'=k\lambda_1$ and $\lambda_2'=k\lambda_2$ is just
a spatial rescaling of the previous system, which consequently 
has the same force $F$ statistics $P(F)$.
As the latter is non-Gaussian for finite $\mu$ it is likewise
so in the asymptotic limit $n_0,\mu\to 0$ with fixed ratio $n_0/\mu$.
For the other limit, $\mu,g\to 0$ with fixed $g^2/\mu$, this
is not the case: rescaling $\mu$ and $g$ to $\mu'=k\mu$ and
$g'=\sqrt{k}g$ does not produce a simple spatial 
rescaling, i.e., a physically equivalent system.

To understand better the meaning of these different limits
it proves instructive to study the spatial
correlation properties of the force. This is the subject of
the next section.

\section{Field-field correlations} 
\label{corr}

Let us consider now, rather than the PDF, the correlation
properties of the force field at two distinct spatial points.
It is instructive to do so first for the original 
unregularized 1D interaction in the finite Poisson 
system of $N$ particles randomly distributed in the 
interval $[-L,L']$. Using Eq.~(\ref{F}),  
we can write
\[
\left<F(x)F(y)\right>=g^2\left<\sum_{i,j}^{1,N} 
\mbox{sgn}(x_i-x) \,\mbox{sgn}(x_j-y) \right>\,.
\]
Using Eq.~(\ref{p-x-n}) it is straightforward to show that
\bea
\label{f-x-y}
&&\left<F(x)F(y)\right>=\frac{N}{L'+L}g^2[L'+L-2|x-y|]\\
&&+\frac{N(N-1)}{(L'+L)^2}
g^2[(L'-L)^2-2(x+y)(L'-L)+4xy]\nonumber
\eea 
From this formula it is clear that, as for the one-point
properties of the unregularized force, this two-point quantity is
ill defined in  the limit $N,L,L'\to \infty$ with
$N/(L'+L)=n_0$. Note that taking the limit symmetrically (i.e. 
with $L'=L$) removes the quadratic and linear divergences 
in the second term, but leaves a linear divergence in the 
first term. Further, in this case, all spatially dependent
terms are in fact finite. This is a crucial point in the
discussion of the renormalisation below, which we now
formulate for convenience using the smooth regularization
procedure used above.
 
Using instead the pair interaction 
defined by Eq~(\ref{f-mu}),
it is straightforward to obtain, using Eq.~(\ref{p-x-n}),
and  taking the limit $N,L,L'\to \infty$ with $N/(L'+L)=n_0$, 
the result that 
\be
G_\mu(x-y)\equiv \left<F_\mu(x)F_\mu(y)\right>=n_0g^2(\mu^{-1}-|x-y|)e^{-\mu|x-y|}
\label{f-x-y-mu}
\ee
which, as one would expect, 
depends only on $|x-y|$. Thus the Jeans' swindle, given by the
$\mu \rightarrow 0$ limit, leaves the expression ill-defined.
Like Eq.~(\ref{f-x-y}) with $L=L'$ the term which diverges
is space-independent, while the space-dependent 
part is finite in the same limit. 
 
It is instructive to give the results for the two-point
properties also in reciprocal space. The Fourier transform (FT)
 with respect to $u=(x-y)$ of the two-point correlation 
function $G_\mu (u)$ corresponds to the power spectrum 
(PS) of the total field, which is thus given by
\be
S_F(k;\mu)=4n_0g^2\frac{k^2}{(k^2+\mu^2)^2}\,.
\label{s-k-mu}
\ee
This result can alternatively be obtained
by calculating directly the FT of $F_\mu(x)$, as
in Appendix \ref{appI}. The integral of the PS of a stochastic field is 
equal, by definition, to its one point variance,
i.e. the integral over $k$ of Eq.~(\ref{s-k-mu})
is equal to $G_\mu(0) = C_2(n_0,g,\mu)$. The divergence
of $G_\mu(0)$ as $\mu \to 0$ corresponds
in $k$-space to the non-integrability of the 
PS ($\propto 1/k^2$) at small $k$, i.e., due
to the ill defined large distance correlation behaviour
\footnote{We note that in any dimension, for the screened interaction,
  at small $k\ll \mu$ (i.e., at scales much larger than the screening
  length $\mu^{-1}$) the PS $S_F(k;\mu)\sim k^2$.  This means that
  $F(x)$ is a so-called {\em superhomogeneous} \cite{glass-like}, or
  {\em hyperuniform} \cite{torquato-hyper}, stochastic field. The main
  properties of this class of stochastic fields are that
  $\int_{-\infty}^{+\infty}dx\, G_\mu(x)=0$, and that the fluctuations
  of the field are sub-poissoniann, i.e., in one dimension the normalized
  fluctuations in a region of size $l$ decrease more rapidly than
  $l^{-1/2}$. For the case of a PS proportional to $k^2$ they decay as
  $\sim l^{-1}$, which is the most rapid possible decay for any proper
  stochastic process in one dimension}. We note that the result
Eq.~(\ref{s-k-mu}) for the PS of the screened gravitational force
field is in fact valid (up to a constant) in any spatial dimension.
In three dimensions, however, the unscreened limit $\sim 1/k^2$ is integrable at
small $k$ (but non-integrable at large $k$). This is an equivalent way
of explaining why the Jeans' swindle (formulated using an exponential
screening) does not work in one dimension while it does in three dimensions.  
Note, however, that the divergence of the variance alone does not
imply in itself that the PDF itself is undefined, which 
we have shown to be the case in one dimension. Indeed the total force 
PDF in three dimensions (i.e. the
Holtzmark distribution) has infinite variance, due however to the
singular behaviour of the 3D pair interaction at vanishing
separation and not to the large scale contributions.

Let us now consider again the renormalization schemes introduced in
Sect.~\ref{ren}. It is clear that in both cases we obtain
\be
G_\mu(x)\to C_2\equiv \left<F^2\right>_{ren}
\label{f-x-y-ren}
\ee
or, equivalently,
\[S_F(k;\mu)\to 2\pi C_2 \delta(k)\,,\]
where $\left<F^2\right>_{ren}$ stands for the (finite) field variance 
after the renormalization procedures. These procedures thus  
keep finite the dominating and diverging contributions 
to Eq.~(\ref{f-x-y}), but at the same time send to zero 
all other subdominant contributions. In particular this means 
that they send to zero all terms depending on the spatial 
argument of the correlation function.
In other words, while the regularization+renormalization procedure, in
both schemes, makes $F$ well defined as a one-point stochastic
quantity, the by-product is to eliminate any space variation of this
field.  The field which remains is a finite acceleration
off-set of the whole system. From a dynamical point of view
this corresponds simply to a translation of the whole system,
and these renormalisation procedures thus erase all information 
about the relative motion of particles. Indeed, in the
spirit of Mach's principle, all relevant physical 
information about the force field is lost in this limit:
``A particle's inertia is due to some (unfortunately unspecified)
interaction of that particle with all the other masses in the
universe; the local standards of nonacceleration are determined by
some average of the motions of all the masses in the universe, [and]
all that matters in mechanics is the relative motion of all the
masses'' \cite{mach}.

These observations naturally lead us to consider the statistics
of the {\it differences} in forces between spatially separated points
in the infinite system limit.

\section{Relative field analysis}
\label{rel-field}

Let us begin again by considering the homogeneous Poisson
distribution defined by Eqs.~(\ref{n}) and (\ref{p-x-n}) 
of $N$ particles in the interval $[-L,L']$ and interacting by the
pair force field (\ref{f}).
Let us fix an arbitrary point, say the origin $x=0$, and 
consider the difference 
of the total field at a point $x$ and the origin:
\be
\delta F(x)\equiv F(x)-F(0)=2g\sum_{i=1}^N[\theta(x_i-x)-\theta(x_i)] \,.
\label{D-F}
\ee 
As the sum is simply, up to a sign depending on $x$, equal to the   
number of particles in the interval $[0,x)$, $\delta F(x)$ is 
manifestly independent of the extremes of the interval $[-L,L']$ and 
therefore remains the same and well defined in the 
limit $N,L,L'\to\infty$ with fixed $N/(L+L')=n_0>0$ taken in any 
arbitrary way. It is thus simple to calculate the one-point PDF 
of $\delta F(x)$, using the
properties of the Poisson distribution:
\be 
P(\delta F;x)=e^{-n_0|x|}\sum_{l=0}^\infty \frac{(n_0|x|)^l}{l!}\delta[\delta
F+2lg(\mbox{sgn}(x))]
\label{P-D-F}
\ee
In other words $\delta F(x)$ can take only (positive or negative depending on the sign of $x$) integer multiple values of $2g$ with a Poisson probability distribution of mean $n_0|x|$.
Proceeding as for the analysis of $P(F)$ in Sec.~\ref{regu},
using the identity Eq.~(\ref{identity}), we can derive 
the characteristic function $\tilde{P}(q;x)=FT_{\delta F}P(\delta F;x)$
(where $FT_{\delta F}$ indicates the FT with respect
to $\delta F$) as 
\[
\tilde{P}(q;x)=\exp \left[{\cal Q}(q;x)\right]
\]
where the cumulant generating function ${\cal Q}(q;x)$ of 
$\delta F(x)$ is given by
\be
{\cal Q}(q;x)=-n_0|x|\left(1-e^{i2gq\, {\rm sgn} (x)}\right)\,.
\label{cumulant-D-F}
\ee
Using the definition Eq.~(\ref{cumu}) we can obtain
the cumulants $\lambda_j(x)$ of $\delta F(x)$ for
all $j \geq 1$ as
\begin{equation}
\label{cumulants-diffPDF}
\lambda_j(x)=[-2g\,{\rm sgn}(x)]^j n_0 |x| \,.
\end{equation}
To calculate the same quantity, but now using the ``smooth''
Jeans' swindle formulated as the $\mu \to 0$ limit 
of the screened gravitational interaction 
Eq.~(\ref{f-mu}), we follow, as in Sec.~\ref{regu},
the Chandrasekhar derivation starting directly
from the definition of the PDF:
\bea
\label{P-D-F2}
&&P_\mu(\delta F;x)= \int..\int_{-L'}^L\prod_{i=1}^N
\left[\frac{dx_i}{L+L'} \right] \times\\ &&\delta\left\{\delta F
-g\sum_{i=1}^N [{\rm sgn}(x_i-x)e^{-\mu|x_i-x|}-{\rm sgn}(x_i)
e^{-\mu|x_i|}]\right\}\,.  \nonumber
\eea 
Following the analogous manipulations and taking the limit $N,L,L'\to\infty$ with fixed $N/(L+L')=n_0>0$ as in Sec.~\ref{regu}, we obtain that the
cumulant generating functional can be written as 
\bea
{\cal Q}_\mu (q;x) &=&-n_0 \int_{-\infty}^{\infty} dy \times \\
&&
\left( 1 - e^{-iqg[{\rm sgn}(y-x)e^{-\mu|y-x|}-{\rm sgn}(y)e^{-\mu|y|}]}
\right)
\nonumber
\eea
and thus the cumulants are given as 
\bea
\label{cumulants-mu}
\lambda^{\mu}_j(x)&=&(-g)^j n_0 
\int_{-\infty}^{\infty} dy \times \\
&&\left[{\rm sgn}(y)e^{-\mu|y|} - {\rm sgn}(y-x)e^{-\mu|y-x|} \right]^j
\nonumber
\eea
First of all we notice that
\[\lambda^{\mu}_1(x)=0\;\;\forall \mu>0\]
It is, furthermore, straightforward to verify that for integer $l$
\[
\lambda^{\mu}_{2l+1}(x)=-\lambda^{\mu}_{2l+1}(-x)\,,
\]
while
\[
\lambda^{\mu}_{2l}(x)=\lambda^{\mu}_{2l}(-x)\,.
\]
Finally one can show that for $j\ge 2$ 
\be 
\lim_{\mu \to 0}\lambda^{\mu}_{j}(x)= \lambda_{j}(x) 
\ee 
i.e., the cumulants of
order $j\ge 2$ converge, for $\mu\to0^+$, to those derived above
in Eq.~(\ref{cumulants-diffPDF}) for
the case without screening, while the average value ($j=1$) instead
vanishes: the Jeans' swindle simply removes the average density, thus
making the average force zero everywhere.
 The associated generating functional is thus \be {\cal
      Q}(q;x)=-n_0|x|\left(1+2igq\, {\rm sgn} (x)-e^{i2gq\, {\rm sgn}
      (x)}\right)\,.
\label{cumulant-D-F-mu}
\ee
Note that this result may be obtained directly from
Eq.~(\ref{P-D-F}) by simply replacing $\ell$ inside
the delta function by $\ell- n_0 |x|$, i.e., by 
simply subtracting by hand the contribution of
the mean density $n_0$ to the difference in the 
force $\delta F(x)$. Note that, conversely, we
can also obtain the initial result (without the
Jeans' swindle) using the second derivation, but
putting $\mu=0$ in Eq.~(\ref{cumulants-mu}) 
before doing the integral. In other words, the 
two limits, (i) the extremes in the 
integral in Eq.~(\ref{cumulants-mu}) and (ii) $\mu\to 0$, cannot be exchanged.
The price to pay for this exchange is the uniform contribution 
coming from the mean density.

It is straightforward also to calculate the two-point 
correlation functions of $\delta F(x)$. 
By using again the joint particle positions PDF of Eq.~(\ref{p-x-n}) 
we can simply evaluate the averages and then take the thermodynamic limit
which is now well defined, finding
\bea
\label{corr-diff-noJS}
G_{\rm diff} (x,y)&\equiv& \left<\delta F(x)\delta F(y)\right>= \\
&&2g^2n_0 [|x| + |y| - |x-y| +2n_0 |x||y|]
\nonumber
\eea
for the unregularized case (i.e. without Jeans' swindle), and
\be
G_{\rm diff} (x,y)
=2g^2n_0
[|x| + |y| - |x-y|]
\label{corr-diff-JS}
\ee
when the Jeans' swindle is used. 
The latter result is
most easily recovered by calculating the correlation
function at finite $\mu$ using the result in
Eq.~(\ref{f-x-y-mu}), and then taking the limit
$\mu \to 0$. The additional quadratic term in 
Eq.~(\ref{corr-diff-noJS}) is simply the contribution
from the non-zero mean density. The interpretation 
of the other terms, common to both expressions,
is very simple: (i) when $x$
and $y$ have different signs, the two intervals 
$[0,x)$ and $[y,0)$ have empty intersection, 
and, as there is no correlation between the
position of particles in a Poisson distribution,
the fluctuations in the variables $\delta F(x)$ 
and $\delta F(y)$ are statistically independent,
and the correlation function [the ``connected'' 
part for Eq.~(\ref{corr-diff-noJS})] is therefore zero; 
(ii) if instead $x$ and $y$ have the same sign, 
the segments $[0,x)$ and $[0,y)$ overlap with 
intersection equal to the shorter of the two 
segments, and thus there is an non-zero
correlation, proportional to the length of
this interval.

\section{Conclusions}
\label{conc}

In conclusion let us consider the implications of our results
for the question of whether an infinite system limit may
be defined for the dynamics of a 1D system of points 
interacting by the 1D version of Newtonian gravity,
when the initial distribution of these points is 
Poissonian. More specifically we wish to consider
the dynamics of particles described by equations of 
motion in one dimension given by\cite{sl1d}
\begin{equation}
\label{sheets-3dexpansion}
\ddot{x}_i +
2H \dot{x}_i= -\frac{g}{a^3}
\lim_{\mu \rightarrow 0} \sum_{j\neq i}
\textrm{sgn}(x_i - x_j) e^{-\mu \vert { x}_i - {x}_j \vert},
\end{equation}
where $a(t)$ is a function describing the expansion of
a 3D universe, and $H=d(\ln a)/dt$ is the corresponding
Hubble expansion rate (and the case $a(t)=1$ describes
the static universe limit). While in the limit $\mu\to 0$ 
these equations are
explicitly well defined for a finite number of particles,
the question is whether they remain well defined when
we consider the usual thermodynamic limit (as defined
above: $L,L' \to \infty$ at constant $n_0=N/L+L'$).
The importance of this limit is that it models the case
of an infinite universe, which is the application of
relevance for these toy models.

We have shown in \cite{sl1d} that this limit may indeed be defined in
the case of an infinite array of particles initially displaced off a
perfect lattice, for a broad class of such displacements. More
specifically we did so by calculating the PDF of the force as defined
on the right-hand side of Eq.~(\ref{sheets-3dexpansion}), and showing
it to be well defined for this class of distributions. In this paper
we have instead shown that the analogous PDF is not well defined for
the case of a homogeneous Poisson particle distribution in the same
limit, and thus that the infinite system limit is not defined in the
same sense. 

In our discussion of other possible regularisations of this
limit, we have shown that the divergence in the total force arises from the
dominant contribution of particles infinitely far away. Because the
1D gravitational force is independent of separation, however, this
component does not contribute to the relative force on any two
particles at a finite distance. As a result, while the force at any
point itself becomes completely undetermined, the difference between forces at two spatially
separate points does not. This means that while Eq.~(\ref{sheets-3dexpansion}) 
is a badly defined equation of motion for each particle $i$,
one can nevertheless write a well defined equation for the
{\it relative} displacements $\delta x_{ij}\equiv x_i - x_j$ 
of two particles $i$ and $j$: 
\begin{equation}
\label{sheets-3dexpansion-diff}
\ddot{\delta x}_{ij} +
2H \dot{\delta x}_{ij}= \lim_{\mu \rightarrow 0} 
\left[ F_\mu (x_i) - F_\mu (x_j) \right]
\end{equation}
where
\begin{equation}
F_\mu (x_i)= -\frac{g}{a^3}
\sum_{k \neq i}
\textrm{sgn}(x_i - x_k) e^{-\mu \vert { x}_i - {x}_k \vert}\,.
\end{equation}
Thus, if we consider the evolution from homogeneous Poissonian initial
conditions, the position of a particle after any finite time 
will always depend on $\mu$, and
diverge with probability $1$ as $\mu \to 0$. On the other hand,
the relative position of any two particles initially at a finite distance
will extrapolate to a finite $\mu$-independent value
in the same limit. In other words, the 
{\it clustering dynamics} of the system ---
entirely characterised by the relative positions 
of particles (e.g. by two or higher point connected correlation 
properties of the density field) --- is well defined. 
In the spirit of Mach's principle, the diverging 
absolute displacement, in an infinite system which 
has intrinsically no centre or preferred point, is 
not, in any case, of physical relevance.

We have shown the above statements to be true strictly only at the
{\it initial time}, i.e., the clustering dynamics is well defined in
an infinite Poisson particle distribution as we have described. That
they remain true as the system evolves away from the initial Poisson
distribution can be most easily verified by considering the evolution
of the density perturbations in $k$-space. It is well known (and
straightforward to show --- see e.g. \cite{peebles}) that, in any
spatial dimension, the evolution of a self-gravitating system leads to
a $k$-independent amplification of the PS of density fluctuations at
small $k$, provided the PS does not vanish faster than $k^4$ at
$k=0$. The small $k$ (i.e. large scales) scaling behaviour of 
the PS of a Poisson particle distribution is thus unchanged
by evolution. As seen above [cf. discussion around Eq.~(\ref{s-k-mu})], 
it is this behaviour which determines the convergence properties 
of the force. These properties therefore 
remain invariant under evolution, guaranteeing that the clustering 
dynamics remains well defined.

The meaning of this limit is that it corresponds to the clustering
dynamics of scales much smaller that $1/\mu$ for the screened pair
interaction, and the limit $\mu\to 0$ sends this upper cut-off 
scale to infinity.
We remark that this limit exists because 
the time scale for evolution of clustering at a given spatial scale
$x>0$  {\it increases} with $x$, i.e. the clustering is what is known as 
``hierarchical'', proceeding from the smallest scale upwards.
It is easy to show heuristically, as follows, that this
corresponds, in general, to a condition on the scale 
dependence of the relative forces which is indeed satisfied
here. The characteristic time scale $t_x$ for evolution of 
a system on a scale $x$ can be estimated as  
\be 
t_x \sim \sqrt{ \frac{x}{|\Delta F (x)| }}
\ee 
where $|\Delta F (x)|$ is the typical relative force on points
at separation $x$. For the case of 1D gravity in an infinite
Poisson distribution, employing the Jeans' swindle as above,  
we have $\left<[F(x)-F(0)] \right>=0$ and therefore we take
$|\Delta F (x)|^2=\left<[F(x)-F(0)]^2 \right>=G_{\rm diff} (x,x)$.
From Eq.~(\ref{corr-diff-JS}) it follows then that  
$t_x \sim x^{1/4}$. Note, however, that if we do not employ
the Jeans swindle, we have instead that the typical force
difference on a scale $x$ is $|\left<[F(x)-F(0)]\right>|=2gn_0 x>0$.
Therefore now $t_x$ is independent of scale, i.e., all scales 
evolve on the same time-scale, which means that the infinite
system limit cannot be defined. This is, indeed, the fundamental
physical reason for the introduction of the Jeans'
swindle: it removes the centre of the system towards
which all points otherwise collapse in a time
independent of the system size.

Finally let us comment on related results given recently in \cite{chavanis}.
This paper derives, using the Jeans' swindle in its usual formulation
as a symmetric ``top-hat'' sum, the PDF of the gravitational force 
in one dimension, defining it taking the equivalent of the renormalised 
weak coupling limit we have discussed. A central point in the paper 
is the observation that in passing from $d=3$ to $d=1$ the statistics
of the gravitational force as characterised by the PDF changes from
the power-law tailed Holtzmark distribution to the Gaussian form
obtained in one dimension Given our results and discussion here, we consider
that there is no basis for giving any significance to this fact:
the Gaussian PDF in one dimension is not obtained in the analagous limit
to that used in three dimensions, and the modified renormalized limit which
gives it has only a trivial physical significance as it leads
to a spatially trivial force field. Further we note  \cite{chandra,libro}
that the ``fat'' (i.e. non-integrable) power-law tail of the Holtzmark
distribution in fact arises from the divergence of the pair 
interaction at small separations, and has nothing to do
with its long range nature. Indeed, even without regulation
of the singularity at small separations, other distributions
in three dimensions (e.g. ``shuffled'' lattices with exclusion regions
around each particle \cite{libro}) have
a Gaussian tail in the gravitational force PDF.
More generally, in fact, as we will discuss
in forthcoming work \cite{sr-lr}, the long-range nature of 
a pair interaction does not lead to divergences of all moments of the force PDF
of order larger than a typical value, leaving the PDF
itself defined even though power law tailed. This is exemplified in the case
we have analysed: from Eq.~(\ref{PDF-cumulants}) we
see that the cumulants of the force field diverge
at any order at the same rate as $\mu \to 0$.

We acknowledge Fran\c{c}ois Sicard for many fruitful discussions
in the context of our collaboration on related projects.
We also thank Bruno Marcos and Bruce Miller for useful 
conversations.

\begin{appendix}
\section{Derivation of the power spectrum of the field $F(x)$}
\label{appI}

The power spectrum $S_F(k;\mu)$ given in Eq.~(\ref{s-k-mu}) can be
also derived directly from its definition: \be
S_F(k;\mu)=\lim_{L,L'\to \infty}\frac{\left<|\tilde
  F_\mu(k;L,L')|^2\right>}{L+L'}\,
\label{def-ps}
\ee
where 
\[\tilde F_\mu(k;L,L')=\int_{-L}^{L'}dx\,F_\mu(x)e^{-ikx}\]
with $F_\mu(x)$ given by Eqs.~(\ref{f-mu}-\ref{F-mu}) and $k$ an
integer multiple of $2\pi/(L+L')$.  In order to evaluate
$F_\mu(k;L,L')$ in a simple way it is useful to notice that the pair
force (\ref{f-mu}) can be derived by a simple derivative (and a
  change of sign) from the pair potential $\phi_\mu (x)=-g e^{-\mu
  |x|}/\mu$ which is a solution of
\begin{equation}
\frac{d^2\phi_\mu (x) }{dx^2}-\mu^2 \phi_{\mu} (x)= 2 g \delta(x) \,.
\end{equation}
The FT (in $[-L,L']$) $\tilde{\Phi} (k;L,L')$ of the total potential
associated with a density field $n(x)$ thus satisfies
$(k^2+\mu^2)\tilde{\Phi} (k;L,L'))=-2g \tilde{n} (k;L,L')$. Now
substituting $|\tilde{F} (k;L,L')|^2=k^2 |\tilde{\Phi} (k;L,L')|^2$ in
Eq.~(\ref{def-ps}), and using the definition of the PS of a 
Poisson point process
\[\lim_{L,L'\to\infty}\frac{\left<\left|\tilde{n} (k;L,L')\right|^2\right>}{L+L'}
= n_0\,,\]
we obtain Eq.~(\ref{s-k-mu}).  

\end{appendix}

\end{document}